\documentclass[a4paper,11pt]{article}
\usepackage{boldline}
\usepackage{cancel}

\usepackage{cancel}
\usepackage{adjustbox}
\usepackage{amsmath}
\usepackage{amssymb}
\usepackage[T1]{fontenc}        
\usepackage{lmodern}    
\usepackage[utf8]{inputenc}     
\usepackage{ifthen}

\usepackage{graphicx}
\graphicspath{ {./images/}} 
\usepackage{adjustbox}

\begin{document}
\title{Remarkable Scale Relation, Approximate SU(5), Fluctuating Lattice}
\author{
H.B.Nielsen, Niels Bohr Institute}
\maketitle
\begin{abstract}
  We discuss a series of 8 energy scales, some of which just speculated by
  ourselves,
  and fit the logarithms of these energies as a straight line versus
  a quantity related to the dimensionalities of action terms in a way to be
  defined
  in the article.

  These terms in the action are  related to the energy
  scales in question. So e.g. the dimensionality of Einstein Hilbert action
  coefficient
  is one related to the Planck scale.
  In fact we suppose in the cases described with quantum field theory, that
  there is for each of
  our energy scales a pair of associated terms in the Lagrangian density,
  one ``kinetic'' and one ``mass- or current'' term. We use for our plotting
  of the energy scales the ratio of the dimensionality of say the
  ``non-kinetic''
  to the dimensionality of the ``kinetic'' one  The explanation for our
  phenomenological finding that the logarithm of the energies depend
  as a straight line on the dimensionality defined integer $q$, we give as
  an ontological - i.e. it really
  exists in nature in our model -``fluctuating lattice''
  with a very broad distribution of say the link size $a$.We take it
  Gaussian in the
  logarithm, $\ln(a)$. A fluctuating lattice is very natural in a theory
  with general
  relativity, since it corresponds to fluctuations in the gauge d.o.f. of
  general relativity.
  
  Intriguing are the lowest ones of our energy scales, being by us not
  described by
  quantum field theory as the other ones, but by actions for single particle
  or single string respectivily, because the string scale fits well with
  hadronic strings, and the particle scale is presumably the mass scale of
  Standard Model group monopoles, a bound state of a couple of which might be
  the dimuon resonance (or statistical fluctuation) found in LHC with mass
  28 GeV.
  
\end{abstract}

\section{Introduction}
Here we seek to develop an idea of there being several energy scales competing
for being  candidates for the fundamental energy scale with the Planck scale
$1.22*10^{19} GeV$ first put forward in our work seeing a unification
with an {\bf approximate}\cite{AppSU5}SU(5)\cite{GG, WGG3, GG2, LD, LRSU5}
without susy,
which gives a
unification scale $5*10^{13}GeV$ (also Senjanovic \cite{Senjanovic} considers
minimal approximate SU(5)) much lower than the Planck scale. In our work
on approximate SU(5) we actually by assuming that at the unification scale
the unified coupling should be just three times as weak(because of assuming
a gauge group being a cross product of three copies of the Standard model
gorup, a group for which we argue in \cite{sr, srcim4,srKorfu14}) as a lattice critical
coupling\cite{crit,LRN,LRNwf,flipped,Polonica,confusionetal, Mizrachi} and a system of scales like we shall gointo details with below,
only using the ones we call in this article ``fermion dip'' (approximated by the
top mass), ``Planck scale'' and the approximate SU(5) ``unified scale'', got
very good fitting of the finestructure constants.

While probably most physicists feel rather sure, that the fundamental scales
for velocity and action are respectively light velocity $c$ and the
action quantum $\hbar$, that also the Newton gravitational coupling $G$
should be the thrird dimensionized quantity to use to build a ``fundamental''
system of units, is somewhat less safe. In fact there are several physical
phenomena and regularities - such as our approximate SU(5) scale
$5*10^{13}GeV$ \cite{AppSU5} or seesaw neutrinoes\cite{seesaw}, which point to
different scales, than the
Planck scale pointed out by the Newton coupling constant $G$. It is the point
of the present article to bring a series of such scales, including some, which
are our own speculations though, into a fit by a straight line of the logarithms
of the energy values pointed to versus a certain dimensionality difference $q$
to be discussed below.
In the simplest cases, such as Planck scale and the scale of masses
of the see-saw neutrinoes, we define dimensionality related quantity $q$ as
the dimensionality
of a to the scale
in question related ratio of Lagrangian term coefficients. In fact we take
a ratio of a
coefficient to a kinetic term to typically a mass term; the mass term we
may exchange
with some  coupling to a current term in stead. That is to say, that we
associate with
each of our ``scales''(except some in single particle or string desciption
ones to be discussed by themselves, and fermion tip scale (also to
be discussed below) ) in the series we consider, a pair of Lagrangian
density terms - or have at least such ones vaguely in mind -, a pair with
one term being a kinetic term and the other one ``non-kinetic''(this is the
mass term or the current coupling term typically).

In the actual cases - of scales- studied either the kinetic or the non-kinetic
term, we shall think of, will be in the usual formulation written without any
dimensionized coefficient, and we shall then just remember to insert an extra
minus sign in counting the dimensionality for the kinetic term, while no such
extra sign is needed when it is  the non-kinetic term, which carry the
dimensionized coefficient.

\subsection{A few works involved with similar things}

Somewhat analogous looking at several scales is found by
Funkhouser\cite{Funkhouser}, and Stojkovic \cite{Stojkovic} looks for
the fundamental scale.

The fluctuating lattice, which is the main model in the present
work, can be considered a development of the idea irregular lattice
such as Vergeles\cite{Vergeles} who works on circumventing the problem of
species doubling. Also J. Brockmann and J. Frank\cite{Brockmann} work with
an irregular lattice, and you may see the lectures by Martin L{\"u}cher
\cite{Lucher}.

A fluctuating lattice was used for difusion equation by
Alexander J. Wagner and Kyle Strand\cite{Wagner}.  
\subsection{A pedagogical example}

The very simmplest, case from which to extend to other cases of
energy scales, is to compare the two scales here called ``see-saw''\cite{seesaw}
and
``scalars'', which are supposed be two energy scales at which
you shall find respectively a lot of Fermion masses (for the ``see-saw'')
and a lot of bosons (for the ``scalars''). The ``see-saw''-scale is the
one supposed to deliver the right handed or Majorana neutrinoes in turn
giving us neutrino oscillations for the usual (left handed) neutrinoes.
The scale``scalars'' should similarly give us a lot of bosons - but it is
purely our invention or speculation of our own (and note that the only true
scalar, the Higgs is not at all at this scale in our  fitting!
The ``scalars'' scale turn out around $4*10^8GeV$, while
Higgs is at 125GeV or 246 GeV.). However,
we add to it the hypotesis, that
there  are more symmetries than we know in the Standard Model, and that
some of them
are broken down spontaneously by means of some of the bosons having masses
in the mass-scale ``scalars'' range and being in some cases tachyonic, so that
assuming that the (``fundamental'') couplings are of order unity the
vacuum expectation values of the fields breaking the symmetries will be
of the order the ``scalars'' scale, too.
Then one may see that weakening of an amplitude (a Feynman diagram say)
breaking one of these symmetries will be of the order
\begin{eqnarray}
  \hbox{Weakening factor }&\approx& \frac{``scalars''}{``see-saw''},\label{sss}
\end{eqnarray}
so that we in this way can connect our fitting to some experimental data
even in the case of our invented scale ``scalars''; we have namely in studying
the spectrum of the quarks and leptons in the Standard Model typically
rather large mass ratios supposedly due to some approximately
conserved symmetries\cite{FN}. Our idea is to identify the typical mass 
ratios, which in some phenomenological model could be due to just
one symmetry breaking, with the ratio of the two scales(\ref{sss}).

Now the prediction from the ``fluctuating lattice''\cite{AppSU5} which is
the name
of the model, which shall below explain our relations.  Let us present
the relatively simple example of the seesaw and the scalars scales:

Compare the Lagrangian densities in quantum field theory for fermions and
bosons (we ignore the interactions and jusy look for the free
Lagarangian density terms, nor for notational dependent factors 1/2, nor the
signs)
\begin{eqnarray}
  {\cal L}_f &=&i \bar{\psi}
  \cancel{\partial}
  \psi + m_f\bar{\psi}\psi\\
  {\cal L}_b &=& g^{\mu\nu}\partial_{\mu}\phi* \partial_{\nu}\phi -m_b^2\phi\phi
\end{eqnarray}
(we here cared only for orders of magnitude)
The important point is,
that the   to be identified with energy scale or masses $m_f$ for the fermions
and $m_b$ for the bosons  come into
the Lagarangian in two different powers (which we denote by the
letter $q$), $m^q$,i.e.
\begin{eqnarray}
  m_f \hbox{ comes to the power } q&=&1\hbox{ i.e. }m_f\\
  \hbox{while } m_b \hbox{ comes to the power} q&=&2 \hbox{ i.e. } m_b^2.
\end{eqnarray}

For dimensional reasons this means, that in a
``fluctuating lattice''\cite{AppSU5}
in which the link size $a$ varies from place to place and has quantum
fluctuations,
we get the averages order of magnitudewise - assuming that couplings, that are
dimensionless are of order unity -
\begin{eqnarray}
  \hbox{Average } m_f &=& <1/a>\label{fm}\\
  \hbox{while Average } m_b&=& \sqrt{<1/a^2>}\label{bm}.
\end{eqnarray}

For narrow distributions in $1/a$ these two expressions are close, but we shall
end up with a distribution in our model having been fitted, which is very broad
indeed and the two different numbers (\ref{fm}) and (\ref{bm}) deviate by a
factor of the order of 251 (which is going to be a ``holy number''in this
article).

Of course there is no fun in getting two points corresponding to two scales
on a line, but of course we can claim just, that our interest in the present
work is the extension of seeing, that plotting the logarithm of the
mass scales - here $log( m_f)$ and $log(m_b)$ - as function of $q$
(= the power) is
our interest; but then it is crucial to truly extend the series of scales
if not to 8 as we do,then at least to three.

\subsection{Outline}
In the next section, number \ref{defq}, we generalize the idea
of the ``power'' $q$ by an extension, that makes it possible to give it
a meaning, provided you relate the scale in question to a ratio of the
coefficients of two terms in the action denotable as ``kinetic'' and
``non-kinetic''(in the case of the Planck scale, gravity, it is rather
all the matter part of the action, which plays the of the role of the
``non-kinetic''
term replacing the mass terms from our example, ``see-saw''and ``scalars'').
In section \ref{maint} we already put forward our main table listing all the
8 scales and their fit, although we still postpone to explain other ones
of the scales but the first - highest- ones, which are describable in term of
coefficients of Lagrangian density terms in a quantum field theory
(namely ``Planck scale'',``see-saw scale''and ``scalars scale'').

In section \ref{ds} we go through the series of scales and explain, what they
mean and their for us important quantities, an effective
``dimesion difference'' $q$ and the energy
of the scale. We seperate the scales into groups and deliver two plots for
the high and the small ones among the scales respectively, but at the end
we also show the full straght line, which fits very well for all our 8 scales.

In section \ref{FGA} we deliver a theoretical argument based on an old idea
of ours for explaining, that there are gauge symmetries in nature\cite{FNN}
which with gravity leads also to the idea
of the fluctuating lattice. This argumentation has basically nothing to do
with our many well
fitting scales a priori, but it rather an independent argument for
a fluctuating lattice.

The conclusion is in section \ref{con} and the outlook means, that one can hope
a much lower fundamental energy scale, e.g. what we call the ``fermion tip''
scale at $10^4 GeV$,than we would be used to think because of expecting
that the most fundamental scaleis the Planck scale.

\section{Definition of a Dimensionality quantity $q$ and Fluctuating
  Lattice Model}
\label{defq}
\subsection{On the variation of actions under $a$, Fluctuations in
  Fluctuating Lattice}

To extract a mass of a particle say you usually need two terms in the action:
a kinetic term (for the particle in question) and the mass term for it.
These two terms should be proportional to the field of the particle, the
mass of which we go for, raised to the same power usually to the second
power. From dimensional arguments one can figure out from the dimensionality
of the constant coefficient $C_i$ to the pure field factors ${\cal L}_i(x)$
to with which power of the lattice constant $a$ this factor ${\cal L}_i(x)$
shall behave as function of the lattice constant $a$. In fact one can use that
the contribution to a little bit of space time to the action $\int C_i
{\cal L}_i(x) d^4x$ must be dimensionless, because the action is dimensionless
in our notation with $\hbar =1$. In fact
\begin{eqnarray}
  \hbox{Since contribution } \hbox{``Bit of action'' }&=&\int C_i{\cal L}_i(X)
  d^4x \hbox{is dimensionless,}\\
  \hbox{and } \#\hbox{ hybercubes in $d^4x$} & \propto& 1/a^4\\
  \hbox{taking the fields in ${\cal L}_i(x)$ to have }dim{field_i}&=&
  \hbox{``sum of dimensions of fields''}\\
  \hbox{and } \# derivatives&=& \hbox{the number of $\partial_{\mu}$ 's}\\
  \hbox{we get } dim_E({\cal L}_i)&=& \# derivatives + dim(fields_i)\\
  \hbox{so with }dim_E(d^4x) &=& -4\\
  dim_E(C_i)&=& -dim_E({\cal L}_i)+4\\
  &=& - dim(fields_i)-\# derivatives +4\nonumber\\
  \hbox{Crudely } C_i &\sim& 1/a^{dim(C_i)}\hbox{(for dimensionless}\\
  &&\hbox{ coupling
      being of order unity)}\nonumber
\end{eqnarray}

Let us be specific of how locally seen the lattice constant $a$ varies
under the fluctuations of it in the fluctuating lattice.

It fluctuates so that in given piece of space time, say an infinitesimal
one $d^4x$ the number of hybercubes goes as $\propto (1/a)^4$,while each cell
four-volume compensate for it by going as $a^4$.

When a field like the field in a gauge theory $A_{\mu}$ has dimension
$[length^{-1}] = [energy]$ one will extract it from the lattice variables
by dividing it explictely by a lattice constant like e.g.
\begin{eqnarray}
  A_{\mu} (x) &\approx& (U(-) -{\bf 1})/a\hbox{(for small deviation of $U(-)$
    from ${\bf 1}$)} ,
  \end{eqnarray}
This is of course in the gauge field case suggested to be natural, because
the field $A_{\mu}$ reminds of a differential quotient then.

If we want a Klein Gordon field, a scalar field $\phi(x)$ in the usual
notation of it having dimension $[energy]$, there are no such ``excuse'',
and if we want to express it in terms of an $a$-independent site
variable $\Phi$, we simple have to by definition take
\begin{eqnarray}
  \phi(x) &=& \frac{\Phi(\cdot)}{a}.
    \end{eqnarray}

When we want to extract e.g. the mass of a scalar field, we compare
coefficients of the two terms
\begin{eqnarray}
  \hbox{``kinetic''} &=& g^{\mu\nu}\partial_{\mu}\phi^{\dagger}* \partial_{\nu}\phi\\
  \hbox{`` non-kinetic''}&=& m^2\phi^{\dagger}\phi
\end{eqnarray}
in the Lagrangian density. The two terms may, say, behave as function of
the lattice constant $a$ as $a^k$ and $a^l$ repspectively and the ratio
of the ``non-kinetic'' to the ``kinetic'' goes as $a^{(l-k)}$, which
is $a^{-2}$ for dimensional reasons in the case of a scalar particle looking
for its mass square. In fact in the usual notation we have $k=0$ and
$l=-2$.

\subsection{Effect of Strong Flutuations on Averages}
Now we are interested averaging this ratio over the fluctuations of the
fluctuating lattice, which we suppose has the Gaussian distribution
in the logarithm of the link size $a$, i.e. we assume that this size $a$ has the
distribution
\begin{eqnarray}
  P(\ln a) d\ln a &=& \frac{1}{\sqrt{2\pi \sigma}}\exp(-
    \frac{(\ln a -\ln a_0)^2}{2\sigma}) d\ln a \label{Gdis}
  \end{eqnarray}
(The more detailed motivation for this assumption (\ref{Gdis}) is given
in subsubsection \ref{mot} just below.)

The main little calculation now is to obtain the average under the fluctuation
of a power $a^p$ of the link size $a$:
\begin{subequations}
\begin{eqnarray}
  <a^p> &=& \frac{\int P(\ln a) a^p d\ln a}{\int P(\ln a) d\ln a}\\
  &=& \int \frac{1}{\sqrt{2\pi \sigma}}\exp(-
  \frac{(\ln a -\ln a_0)^2}{2\sigma}) a^p d\ln a\\
   &=&\int \frac{1}{\sqrt{2\pi \sigma}}\exp(-
  \frac{(\ln a -\ln a_0)^2}{2\sigma}+p\ln a)  d\ln a\\
  &=&  \int \frac{1}{\sqrt{2\pi \sigma}}\exp(-
  \frac{(\ln a -\ln a_0 +p\sigma)^2 -(-\ln a_0+p\sigma)^2 +(\ln a_0)^2}
       {2\sigma})  d\ln a\nonumber\\
       &=& \int \frac{1}{\sqrt{2\pi \sigma}}\exp(-
  \frac{(\ln a -\ln a_0 +p\sigma)^2 + 2\ln a_0 *p\sigma-(p\sigma)^2 }
       {2\sigma})  d\ln a\\
       &=&a_0^p * \exp(-p^2*\sigma/2)\\
       &=& (a_0\exp(-p/2*\sigma))^p
\end{eqnarray}
\end{subequations}
This little calculation may be interpreted to say, that the effect of the
fluctuation in $\ln a$ with the spread $\sigma$ leads to replacing the
a priori lattice link size $a_0 $ as
\begin{eqnarray}
  a_0 &\rightarrow & a_0\exp(-p/2*\sigma)\\
  \hbox{or }1/a_0 &\rightarrow &1/a_0 *\exp(p/2*\sigma)\\
  \hbox{for looking for } a^p & or & (1/a)^{-p}
\end{eqnarray}

Let us say, we want to see the effect of change
for action 
wherein the 
``non-kinetic'' term has a coefficient with dimention $[mass^q]$ higher
than the ``kinetic'' one. So from dimensional reasons the ``non-kinetic''
term has to behave with a factor $(1/a)^q$ higher than the ``kinetic'' one.
This means that this ratio is $a^{-q}$ and the original lattice
size $a_0$ will for the purpose of this mass be replaced
\begin{subequations}
\begin{eqnarray}
  a_0 &\rightarrow& a_0\exp(q/2*\sigma)\\
  \hbox{or } m=a_0^{-1} &\rightarrow& a_0^{-1} \exp(-q/2 \sigma)
  = m\exp(-q/2*\sigma)\hbox{(in the cases like (\ref{fm}) and (\ref{bm}).)}
  \nonumber
\end{eqnarray}
\end{subequations}

We can of course only expect such formulas to work for particles, which are not
mass-protected, such as the see-saw right handed or Majorana neutrinoes,
so the fermion mass scale given by the mass power $q=1$ is by us considered
the see-saw-neutrino scale and denoted by such in our table. Similarly
for the $q=2$ scale, which we call ``scalar scale'', we speculate, that many
bosons have masses of that order of magnitude, and associated with that
also some expectation values of boson fields breaking presumably various
symmetries. In fact we believe we can include the Planck scale with an
effective $q=-2$. For the gravity the Einstein-Hilbert action is of course
to be considered a ``kinetic'' term, and we may consider the
matter action as essentially taking the role of the mass term in the
analogy with our first examples. But now the coefficient to the ``kinetic''
Einstein-Hilbert is wellknown to be of dimension $[1/(8\pi G)] = [mass^2]
=[energy^2]$. This would correspond to the excess of the
``non-kinetic''-coefficient
over the ``kinetic'' one to be $[m^q] = [m^{-2}]$ i.e. $q=-2$.

\subsubsection{Motivation for the Gaussian Distribution in the Logarithm}
\label{mot}
We would tend to argue that such statistical distribution as (\ref{Gdis})
is really a very often appering one. Although it were actually known
before 1931 it is refered to at least by Kalecki\cite{Kalecki} as the
``Gibrat distribtuion'' derivable from ``Gibrat's
{\em loi de l*effet proportionnel}, which means that the distribtuion has
modified by some random independent proportionate changes, as we use just below.
Pestieau and Possen\cite{Pestieau} use it studying income distribution.
The here proposed Lognormal distribtuion is also sometimes called
Galton ditribtuion or Cobb-Douglas, see\cite{Johnson}. 

In our case the argumentation for this distribution of a Gaussian in the
Logarithm is even more suggestive considering the section \ref{FGA} below.

In fact we put below in section \ref{FGA} the idea that the gauge of gravity
- taken as ontologically existing being represented by a lattice - is
fluctuating over the results of a huge set of possible gauge transformations
( = reparametrisations), and that we should expect this lattice to be
in first approximation distributed with the Haar-measure distribution
in as far as such a Haar measure exists. But whatever the properties of the
reparametrization group might be, it at least contains scaling as a subgroup.
The Haar measure for the group of scalings is a flat distribution in the
{\bf logarithm} of the scaling factor.We now want to suppose that the
distribution of say the size of the links in the fluctuating lattice
should be approximately so that, if one starts from one link size and
consider all
link sizes obtained by transforming the whole lattice with all scale
transformations weighted with the Haar measure (for the scaling group,
hoped to be the same as for all reparametrizations), one should get the link
distribution proposed. Of course with the true Haar measure for the scaling
group one would get the distribtuion (\ref{Gdis}) with the spread $\sigma$
being infinite. We consider it thus that it is just approximation that we
take instead a ``very large'' $\sigma$. That we take this very flat
distribtuion to be just a Gaussian distribtuion, is the speculation, that
it will likely in some hoped for more detailed theory for it in the future
come out as composed of several fluctuation composed on to of each other
in an approximate scale invariant way. Scaling in successive steps
will mean adding in the logarithm and thus leads to precisely a Gussian
in the logarithm, as we have chosen.

\subsubsection{``unification scale''}
For pure Yang Mills you have no mass term, but could possibly imagine
splitting the Yang Mills action up into some terms with derivatives
being the genuine kinetic terms while others could be thought of
as ``non-kinetic'', but in any case there are no dimensionized quantities
in the action, and thus in the just put forward scheme we should imagnine
that it is the very $a_0$ the a priori lattice scale, which should
correspond to the scale related to the Yang Mills acton. But in the philosophy
of at least an approximate SU(5), we should take the approximate unification
scale giving us $a_0$. In other words the effective $q$ for the unification
scale should be $q=0$.  

We have now seen for the four uppermost of the  energy scales, which I consider,
that they can be classified  by an ``effective mass term dimension'' $q$.

\subsubsection{Postponing three scales}

We shall postpone till below, when discussing these scales, to tell how we
can extend the concept of the ``effective mass term dimension'' $q$
to three more speculated energy scales, which we shall call:
``Fermion tip''(obtained by extrapolating in the appropirate way the
spectrum of quark and leptons in the Standard Model), ``monopole scale''
(from the speculation that there are after all monoples, say for QCD, and
that a bound state of them were found in LHC\cite{dimuon} as a dimuon
resonance of mass
28 GeV), and ``string scale'' (which very mysterious in a way points to the
very old first studied strings being hadrons.).

\section{Our Main tabel}
\label{maint}
But let us to put up the main point of our article so soon as
possible present immediately the table of all the scales, we believe
we can fit in our scheme and then explain and define the scales
below:

\begin{adjustbox}{width =\textwidth}
\begin{tabular}{|c|c|c|c|c|}
  \hline
  Name&[Coefficient]&``Measured'' value&Text ref.&\\
  comming from&Eff. $q$ in $m^q$ term& Our Fitted value&Lagangian dens.&\\
  status&&&&\\
  \hline
  \hline
  Planck scale&$[mass^{2}]$ in kin.t.&$1.22*10^{19}GeV$&(\ref{Planckenergy})&\\
  Gavitational $G$&q=-2&$2.44*10^{18}GeV$&$\frac{R}{2\kappa} $&\\
  wellknown&&&$\kappa= 8\pi G $&\\
  \hline
  Redused Planck&$[mass^{2}]$ in kin.t.&$2.43*10^{18}GeV$&(\ref{ReducedPlanck})&\\
  Gravitational $8\pi G$&q=-2&$2.44*10^{18}GeV$&$\frac{R}{2\kappa}$&\\
  wellknown&&&$\kappa=8\pi G$&\\
  \hline\hline
  Minimal $SU(5)$&$[1]$&$5.3*10^{13}GeV$&(\ref{unification})&\\
  fine structure const.s $\alpha_i$&q=0&$3.91*10^{13}GeV$&
  $\frac{F^2}{16\pi \alpha}$&\\
  only approximate&&&$F_{\mu\nu}=\partial_{\mu}A_{\nu}-\partial_{\nu}A_{\mu}$&\\
  \hline
  Susy $SU(5)$&$[1]$&$10^{16}GeV$&(\ref{susy})&\\
  fine structure const.s&q=0&$3.91*10^{13}GeV$&$\frac{F^2}{16\pi \alpha}$&\\
  works&&&$F_{\mu\nu}=\partial_{\mu}A_{\nu}-\partial_{\nu}A_{\mu}$&\\
  \hline
  \hline
  Inflation $H$&$[1] ?$&$10^{14}GeV$&(\ref{inflationH})&\\
  CMB, cosmology&q=0?&$3.91*10^{13}GeV$&$\lambda \phi^4$&\\
  ``typical'' number&&&$V=\lambda \phi^4$&\\
  \hline
  Inflation $V^{1/4}$ &concistence ?&$10^{16}GeV$  &(\ref{inflationV})&\\
  CMB, cosmology&q=-1?&
  $9.96*10^{15}GeV$&consistency&\\
  ``typical''&&&$V=\lambda \phi^4$?&\\
  \hline
  \hline
  See-saw& $[mass] in \; non-kin.$ &$10^{11} GeV$& (\ref{seesaw})&\\
  Neutrino oscillations& q=1& $1.56*10^{11}GeV$&$m_R\bar{\psi}\psi$&\\
  modeldependent& &&$m_R$ right hand mass&\\
  \hline\hline
  Scalars&$[mass^2] in\; non-kin.$&$\frac{seesaw}{44\; to\; 560}$&(\ref{s43}, \ref{s152})&\\
&&&&\\
  small hierarchy& q=2&$\frac{1.56*10^{11}GeV}{250}$& $m_{sc}^2|\phi|^2$&\\
  invented by me&&&breaking $\frac{seesaw}{scalars}$&\\
  \hline\hline
  Fermion tip& $``[mass^4] in\; non-kin.''$& $10^4 GeV$&(\ref{mnml})&\\
  fermion masses& q=4& $10^4GeV$& ``1''&\\
  extrapolation &&&quadrat fit&\\
  \hline\hline
  Monopole& $``[mass^5] in\; non-kin.''$& $28 GeV$&(\ref{monopole})&\\
  dimuon 28 GeV& q=5 & $40 GeV$&$m_{monopol}\int ds$&\\
  invented&&&$S \propto a$&\\
  \hline\hline
  String $1/\alpha'$ & $``[mass^6] in\; non-kin.''$& $1 Gev$& (\ref{alphaprim})& \\
  hadrons& q=6& $0.16 GeV$&Nambu Goto&\\
  intriguing &&&$S\propto a^2$&\\
  \hline
  String $T_{hagedorn}$& $`[mass^6]$ in\; non-kin.''& $0.170GeV$& (\ref{Hagedorn})&\\
  hadrons&q=6& $0.16 GeV$ & Nambu Goto&\\
  intriguing&&&&\\
  \hline\hline
\end{tabular}
\end{adjustbox}

\subsection{Explanation of the Table}

The uppermost block in the table delimeted by double
horizontal lines is the explanation of the columns, and each of the
successive such blocks delimeted by double lines represent one energy scale
proposed. There are eight such energy scales described. The single horizontal
lines seperate
slightly different versions of the energy scale described in the block
delimited by the double lines, typically only deviations by a number of
order unity, so these multiplication of the same scale a couple of times is
usually not of any significant interest, and may be ignored.
(In the case of the ``inflation scale'' the different formulation though
deviate by morethan an order of magnitude, and the useof susy for unifying
SU(5) scale deviate also more fromthe minimal SU(5) scale,and our model fit
best the no susy.)

The table is in four columns, but in each of these four columns we have
put two
to three different items for each scale.

\subsubsection{Content in the different coloumns}

\begin{itemize}

\item{\bf The first column (from left)} contains the three items:
\begin{itemize}
\item{1.} A name, we just acsribe to the scale in question.

\item{2.} An allusion to, from which data the energy scale number is determined.

\item{3.} What we call ``status'', an estimation of how good the story
  of the scale in question is.
\end{itemize}

\item{In {\bf the second column} the items are:}
\begin{itemize}
\item{ 1.} $[coefficient]$: It is the dimensionality of the coefficient
  to a term
in the Lagrangian relevant for the interaction giving the scale in question.
In the table is added either ``ìn kin.t.'' or ``in non-kin.'', meaning
respectively, that the term with the coefficient of the dimension given
is the kinetic term or the non-kinetic term repectively.

\item{2.} In next line inside this column 2 is written the quantity $q$
  and its value
for the scale in question, or some effective value for this $q$.
For non-kin. $q=dim_{energy}(coefficient)$, while for ``kin.t.'' it is
$q=-dim_{energy}(coefficient)$ because we have in all of cases (by accident),
that the other
coefficient is dimensionless.
\end{itemize}

\item{Then in the {\bf third column}:} comes the numbers, the energy scale.
\begin{itemize}

 \item{1.} In the top line inside the blocks comes the experimental or rather
   best theoretical estimate from the experimental data.(the type of data was
   mentioned
in second line in column 1).

\item{2.} In the next line we have put the fit to the straight line of the
  logarith of
  the energy versus $q$. It is given by the fitting formula:
  \begin{eqnarray}
    \hbox{``fitted value'' }&=& 10^4GeV*(250)^{4-q}\\
    &=&3.91*10^{13}GeV*250^{-q}. 
    \end{eqnarray}
\end{itemize}

\item{In {\bf fourth column}:}
  \begin{itemize}
  \item{1.} In first line is a reference to the formula in the text
    representing the decicion as to, what value to take for the scale in
    question. (Usually at best trustable to orderof magnitude.) 
  \item{2.} In the second line is the Lagrangian density used in determining
    the dimension of its coefficient, $[coefficiient]$.
  \item{3.} In the third line we put some formula or remark supposed to
    make it recognizable, what the Lagrangian density in the line 2
    means.
    \end{itemize}
\end{itemize}

Note in generel that the main point of our paper is that the {\bf two numbers
in the third column agree} for most of the scales. The agreement for the
susy unification is though not so impressive,
while the minimal SU(5) unification and the inflation Hubble constant $H$,
seems to fit better, but the associated $V^{1/4}$ we only get into our scheme
by using its relation to the infaltion Hubble Lemaitre expansion in the
inflation time by LFRW relation. Therefore we wrote ``consistency'' for this
$V^{1/4}$ case..

\section{Discussions of the Several Different Scales}
\label{ds}
\subsection{The Four Hughest and Simplest to Discuss Scales}
\subsubsection{{\bf Planck Scale}}
  The Planck energy scale is the scale defined by means of the
  Newton gravitational constant $G$,and conventionally we define
 \begin{eqnarray}
   \hbox{Planck energy=}  E_P&=& \sqrt{\frac{\hbar c^5}{G}}
   = 1.9561*10^9 J\\
   &=&  1.22*10^{19} GeV,\label{Planckenergy}
 \end{eqnarray}
 but since the quantity that occurs in the Einstein-Hilbert-action
 $S_{EH}=\int\frac{1}{2\kappa}R\sqrt{-g}d^4x$
 is in fact $\kappa = 8\pi G$, it would be indeed very natural to instead
 use the so called ``reduced Planck energy''
 \begin{eqnarray}
   \hbox{Reduced Pl. energy=} E_P/\sqrt{8\pi}&=&3.9019*10^8 J\\
   &=&2.43*10^{18}GeV. \label{ReducedPlanck}
 \end{eqnarray}
 The Einstein-Hilbert action, we just saw, had a coefficient of dimension
 $[1/(2\kappa)] = [GeV^2] = [mass^2]$ and since the scalar curvature $R$
 being of the form of two dervivatives acting on the metric tensor $g_{\mu\nu}$
 it is of course to be considered a ``kinetic'' term. There is of course no
 true mass term since the graviton shall be massless, but we may then claim
 to compare the Einstein-Hilbert-term with some or all the mattter action,
 which contains $g^{\mu\nu}$ e.g.. In any case these matter terms have no
 dimensionized coefficients unless they are involved in some of the other
 scales, and we consider them effectively without any for the Planck scale
 relevant dimensionized coefficient. Thus the ratio of the kinetic term
 compared to the non-kinetic one is $[mass^2]$. Thus counted the opposite way,
 i.e. the ratio of the non-kinetic to the kinetic term coefficient has the
 dimension $[mass^q]=[mass^{-2}]$ and so $q=-2$ for the Planck energy scale.

 \subsubsection{{\bf (approximate) Unification scale}} The Lagrangian
 density term,
  which we associate with this ``unification scale'' is the Yang Mills
  Lagrangian density, which has the dimensionality of energy to the
  fourth power (it happens that because this dimension just
  cancels that of the measure $d^4x$ in four dimensions, the coefficient,
  essentially   the inverse fine structure constants, are dimensionless.
  In fact it is wellknown that the finestructure constants are dimensionless,
  and thus if we consider $F_{mu\nu}^2$ a kinetic term and there is no
  dimensionized candidate for a term to compare with, it is clear that ratio
  of the
  coefficients will be just of zero dimensionality.
  The candidates for non-kinetic term would be either some other part
  of the $F_{\mu\nu}^2$ term than the genuine kinetic term
  with two derovatives or the current-gauge-boson coupling term,
  like $j^{\mu}(x)A_{\mu}(x)$. 
  So $q=0$.
  
  This means then, that there will come no factor due to the fluctuation, and
  the scale of the lattice will be the $a_0$ itself. But now by assuming that
  the {\bf approximate} meeting of the running couplings predicted by SU(5)
  would mean, that a scale {\bf is} pointed out, namely the approximate
  unification scale. We even pointed out in \cite{AppSU5}, that except for
  a factor 3 the deviation of the measured fine strucuture constants from
  exact $SU(5)$ could be interpreted as a lattice quantum correction.
  That is to say we proposed a lattice plaquette action which in the
  classical approximation gave just SU(5) relations between the couplings
  for a Standard Model Group lattice model. We even mannaged using the
  assumption the gauge couplings being critical \cite{AppSU5, Ryzhikh, LR} and
  some rudimentary
  version of the present article scale fitting using what we call
  AntiGUT
  \cite{Volovik, Picek, LR, Ryzhikh, RDrel1, RDrel2,RDrel3, RDrel4,Don93,
    Don137 } to essentially obtain the three
  Standard Model fine structure constants.
  In our ``Approximate $SU(5)$, Fine structure constants''\cite{AppSU5} we
  fit the
  replacement of unification to be from the three measured finestructure
  constants
  \begin{eqnarray}
    \mu_U &=& 5.13*10^{13}GeV,\label{unification}
    \end{eqnarray}
  but we could crudely have gotten this value just by looking at the plot
  of the running inverse fine structure constants as function of the logarithm
  of the energy scale $\mu$ and ask for where there is a neck, where the
  three inverse fine structure constants are closest to each other.

  By just looking roughly on the running coupling constant plot for the
  energy scale at which the ``unify the best'' we get a very similar
  value as the one by our quantum corrections multiplied by three
  value (\ref{unification})

  With susy assumed a number more like (see e.g. \cite{Thoren})
  \begin{eqnarray}
    M_U &=& 10^{16}GeV \label{susy}
  \end{eqnarray}
  is achieved. This $10^{16}GeV$ fits worse in our final fit, and we could
  in that sense
  say, that our model disfavours susy.

\subsubsection{\bf  See-saw Scale}

We collected a few references for estimates of the typical
mass scale for the right handed neutrinoes causing the non-zero mass
differences with their neutrino oscillations for the observed (left handed)
neutrinoes.

Some such quite model dependent values for the right handed neutrino
masses are given in the table \ref{rhs}
\begin{table}
  \label{rhs}
    \begin{tabular}{|c|c|c|c|}
    \hline
    Name & Scale& Quality&\\
    ``True'' Takanishi et al.&$1.0*10^6 \;  to \; 7.8*10^9GeV$&
    somewhat serious&\cite{Takanishi,Taknishi2}\\
    \hline
    Takanishi's and our model &$1.2*10^{15}GeV$&low&\cite{Takanishi}\\
    \hline
    Steven Kings Model&$3.9*10^{10}GeV$&serious (susy)&\cite{King}\\
    \hline
    Simple statistical estimate&$1.4*10^9 GeV$&  a bit&\\
    \hline
    Mohapatra's guess& $10^{14}GeV$ to $10^{15}$& inaccurate&\cite{Mohapatra}\\
    \hline
    Grimus and Lavoura & $\sim \; 10^{11}GeV$& serious&\cite{Grimus}\\
    \hline
    Davidson and Ibarra Bound & $\ge 10^9 GeV$ & serious&\cite{DavidsonI} \\
    \hline
    \end{tabular}
    \caption{In this table we collect some numbers mentioned for the
      right handed neutrinoes that could be connected with the known
      neutrino oscillations , and mostly also to creation of baryon assymetry.
      }
      \end{table}
    The most suggestive value seems to be
    \begin{eqnarray}
      \mu_{see-saw} &=& 10^{11} GeV.\label{seesaw}
      \end{eqnarray}

    Since the the mass term for a feremion - such as a see-saw ``right handed''
    neutrino - is wellknown to have a coefficient deviating from
    that of the corresponding kinetic term by a ratio having the
    dimensionality $[mass]$ it is trivial that for this see-saw scale
    $q=1$. (We saw it already in subsection of the introduction.) 

    \subsubsection{The Scalar scale}

    This ``scalar scale'' represents the speculative assumption, that at
    some scale you will find a lot of scalar bosons with order of magnitudewise
    the same mass. It should basically be the majority of scalars existing.
    (But remember that  the wellknow Higgs falls completely outside our scheme
    for some reason of it being fine tuned in mass presumably. It fails
    our model.)
    Since for a scalar the dimensional ratio of the coefficient of the mass
    term
    to that of the kinetic term is $[mass^2]$ of course such a scale of scalars
    should correspond
    to the scale
    with $q=2$.(We also saw that already in a subsectionof the introduction.)

    We naturally suppose, that assuming coupling constant of order unity a
    series of vacuum expectation values should have the same orders of
    magnitude.

    If this is so, these vacuum expectation values could cause breakings
    of symmetries, that might be found at higher scales. In fact we suggest
    that such expectation values should cause the phenomenon known as the
    little hierarchy problem:

    The ratios of different fermion masses in the Standard Model are offen
    rather high numbers. This would be a problem, if one assumes, that all
    the variuous coupling constants are of order unity. If one, however, have
    one or preferably several {\bf weakly} broken symmetries, which
    distinguishes
    in charges some right and left handed components of the Standard Model
    chiral fermion fields, then such an occurence of large mass ratios
    among the Standard Model fermions would easily and naturally occur.
    This is what is called Froggatt-Nielsen mechanism.\cite{FN}

    But there is still a large number problem in as far as one may still
    ask, could it be natural that the symmetry breakings were {\bf weak}?
    In the present paper we shall answer this question by saying, that we
    get the
    large ratios because the scalar scale is a large number lower in
    energy than the see-saw scale. In this way we refer the problem of the
    big numbers to a common big number giving also the big ratios of our
    different energy scales.

    It is namely so, that the typical Feynman diagrams for giving masses
    to Standard model fermions via breaking of some symmetries will be series
    vacuum expectation values of the boson fields interspaced with propagators
    for fermions. But according to our philosophy the majority of fermions
    have masses of the order of the ``see-saw'' scale, and the majority of
    vacuum expectation values breaking some of the symmetries have expectation
    values of the order of the ``scalar'' scale. So the breaking become
    {\bf weak} by a
    ratio of the scalar scale energy over the see-saw scale energy. This is
    according to our fit of the order of 250. So we predict, that the typical
    ratios of Standard model fermion masses shall be of the order of
    a few factors
    of a number of this order of 250. This is indeed close to being true, in
    as far as the successive charged leptons have ratios not far from 250. 

    To get a number to give as a phenomenological value for the big number(s)
    explaining the small hierarchy problem we looked for some quark mass ratios
    being about equal so as to having hopefully found the ratio comming
    by breaking one of the symmetries supposedly weakly broken
    several times. (But likely one would find better estimates
    by help of the litterature \cite{Takanishi, qarketcmasses, Ferrucchio})
    This search led to the chain of quarks:
\begin{subequations}
    \begin{eqnarray}
      top &with \; mass=& 172 GeV\label{a}\\
      \hbox{Ratio } top/bottom &=&   41.1\\
      bottom &with \; mass= & 4.18 GeV\\
      \hbox{Ratio } bottom/strange &=& 44.9\\
      strange&with \; mass=& 93 MeV\\
      \hbox{Ratio } strange/up &=&42.2\\
      up &with \; mass=& 2.2 MeV.\label{b}
    \end{eqnarray}
    \end{subequations}
    Here obviously the same ratio about 43 occurs three times. So
    assuming the Yukawa-coupling $g_y$  in the mention daigram part to be $1$
    we get the scale-ratio:
    \begin{eqnarray}
      \frac{``see-saw''}{``scalars''} &=& 43\label{s43}\\
      \hbox{but with $g_y=\sqrt{4\pi}$ then: }
      \frac{``see-saw''}{``scalars''}&=&
   43*3.54=152\label{s152}   
      \end{eqnarray}

    \subsubsection{Straight line relation for the ``Simplest Upper Four''}
\begin{figure}
      \includegraphics[scale=0.65]{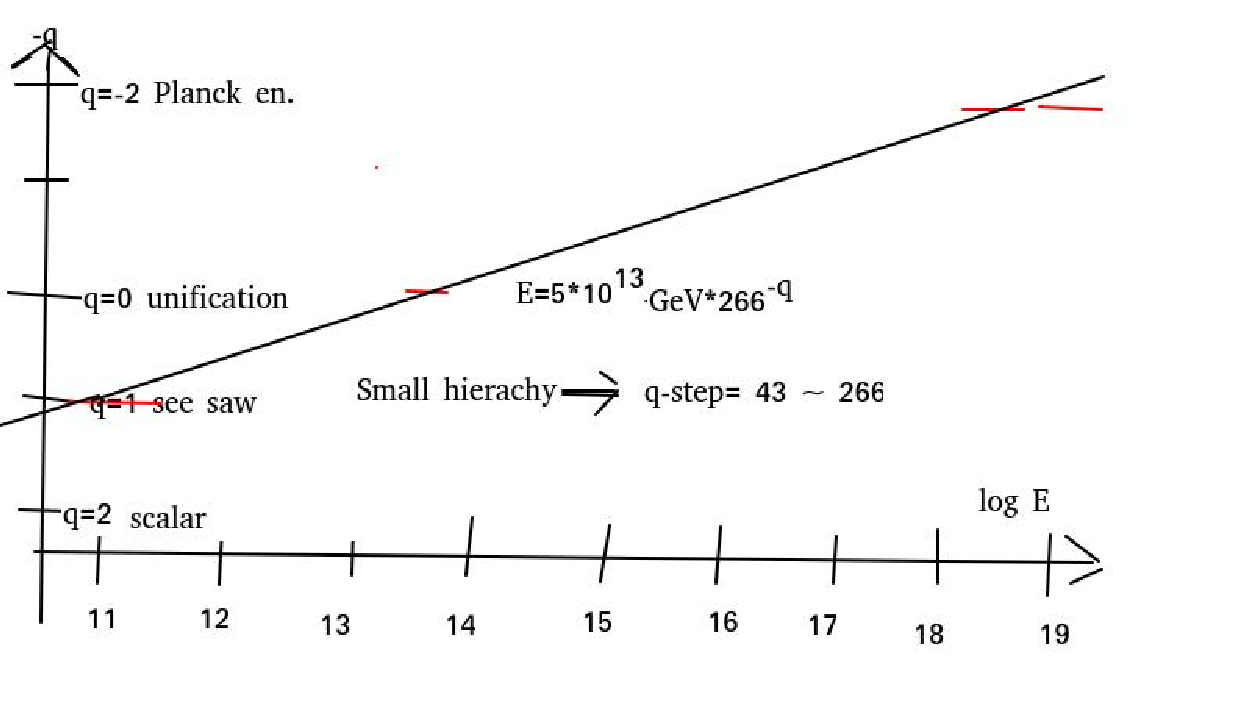}
      \caption{\label{plotf4}The plot of the dimension-of-coefficient-related
        $q$ (actually more negative $q$ goes upwards) versus the 10-logarithm
        of the energy scale measured in GeV (we took the $log_{10}$ rather than
        $\ln$, because we have given the energy scales in powers of ten; so it
        is easier to translate to the energy scales these $log_{10}$ numbers).
        We did not present the point for ``scalar'' directly, because the
        small hierarchy mass ratios rather give us the ratio of the
        see-saw scale energy over the scalar one, and this ratio
        is simply the step-factor in energy scale per lowering of $q$ by
        one unit. The fit for this lowering factor from the three
        scales presented is 266. This is indeed of the order of the
        small hierarchy mass ratios. We found a chain of four quuarkmasses,
        t,b,s,and u, having rather closly the ratio 43 in mass successively
        (see (\ref{a}-\ref{b})).
        This 43 deviates from the 266 by less than a factor $4\pi$ say, in as
        far as it is only deviating by 6.2. Below in the article we shall
        extend this plot to the scales which have a more complicated
        discussion.)}
    \end{figure}

    On the figure \ref{plotf4} we see that indeed the four simplest to asign
    $q$-values for uppermost four scales lie wonderfully on a straight line
    to the only order of magnitude accuracy, which we can hope for.
    The overall energy scale of the energy scales is a dimensionfull number and
    there is of course no way to hope for the theory to predict that, but the
    slope of the straingt line fitting, which corresponds to that for each
    lowering of $q$ by one unit, the energy scale goes up by a factor 266,
    represent a pure number, that a priori could be hoped to be predicted from
    some theory. It is so to speak a ``holy number'' like say the fine
    structure constants.
    
    \subsection{The more complicated scales}
    The remaining four scales we have left here to the end because we consider
    them complicated in a couple of ways: The ``infaltion'' scale is
    a cosmological scale rather than a scale of physical laws, so it 
   might
    at the end not truly belong in our scheme ? The other three scales are
    complicated once you start from the field theory defined scales
    studied at first, because we for these scales do not use a
    field theory description, but rather use a descripton of single strings
    and single monoles,and thus the factor we already talked about, that the
    number of lattice hypercubes in a piece of space time varies
    as $1/a^4$, is no longer relevant for the single particle descriptions.
    
    Finally the ``Fermion tip'' scale involves very speculative extra
    assumtons, but to make up for it is associated with a reasonable
    fit of the density of fermion masses on an $\ln m$-axis.
    
    \subsubsection{The Inflation scale}
    Thinking of the inflation rate $H_{inflation}$ as representing an expansion
    of both space and the lattice somehow following the space expansion it
    would mean a big number being present in the lattice expansion, if during
    one link length step the space the world expanded faster than by of order
    unity. So in the philosophy of taking the couplings of order unity at the
    fundamental level, we should have no faster expansion than that, but
    presumably have just that rate of expansion order of magnitudewise. That
    is to say we expect
    \begin{eqnarray}
      H_{infation} & \approx& <1/a> \approx 1/a_0.
      \end{eqnarray}
    The last argumentation here is that, since we have not found any reason why
    the $a$ should be biased, except for the bias of being looked at in a
    quantum field theory in which one in a fixed continuum region
    $d^4x$ sees a number of hypercubes proportional to $1/a^4$. This is just
    the same bias as for the unification scale, and thus we predict that the
    $H_{inflation}$ shall be similar in order of magnitude to the unification
    scale. That is to say we should assign to the inflation rate scale the
    effective $q$-value
    \begin{eqnarray}
      q|_{inflation\; H \; scale} &=& 0.
      \end{eqnarray}


    We could also ask for the potential $V$ for the inflaton field during
    the inflaton era, or rather consider its fourth root $V^{1/4}$ an energy
    scale. Now, however, this $V$ during inflation is related to the
    inflation rate $H_{inflation}$ supposedly of order $1/a_0$ by the FLRW
    equation. In fact we have
    \begin{subequations}
    \begin{eqnarray}
      \left ( \frac{\dot{a}}{a} \right )^2+ \frac{k}{a^2}-\frac{\Lambda}{3}
      &=& \frac{\kappa}{3}\rho\\
      \hbox{or with } \Lambda &=& \kappa* V\\
      \hbox{and dropping } k \hbox{ and } \rho &=&0\\
      H^2 &\approx& \kappa * V \approx \frac{V}{E_{pl}^2}\\
      \hbox{So in our fit with } H_{inflation}&\approx & 1/a_0\\
      E_{Pl}&\approx& 1/a_0*``stepfactor''^2\\
      V^{1/4} &\approx& 1/a_0 *``stepfactor''^{-q_{V^{1/4}}}\\
      \hbox{we must have } q_{V^{1/4}} &=& -1.
    \end{eqnarray}
    \end{subequations}
    
    Here $``stepfactor''$ was the factor by which the scale goes up each
    time the $q$-value goes down with one unit. We found this ``holy number''
    $``stepfactor'' = 251$.
    

    Indeed the $q=0$ meaning that the Hubble-Lemairtre expansion at
    inflation time should be of same order as the unification scale
    (using approximate SU(5)) is agreeing with that according to 
Engquist\cite{Enquist} it is 
is $10^{14} GeV$, very close to the one for ``unification scale''.

\begin{center}
  {\bf The Inflation Hubble-Lemaitre Constant (at end of inflation)}
  \end{center}

\begin{itemize}
\item{\bf Liddle}
As a scale to represent the energy scale for the inflation period of the
universe we may - perhaps a bit biased - choose the Hubble-Lemaitre expansion
at the end of the inflation: $H_{end}$ which according to Andrew Liddle
\cite{anintr} takes values like

\begin{eqnarray}
  \frac{H_{end}}{m_{Pl}}&=&7.5*10^{-5}
  \left ( 1+\frac{240}{\alpha}\right )^{-\frac{2+\alpha}{4}}\\
  &=&
  6.2*10^{-7}\;
  for \;
  \alpha =2\\
  \hbox{or }  &=&1.6*10^
      {-7}\; for\;  \alpha=4
  \end{eqnarray}

Using the values $1.22 *10^{19}GeV$ or $2.43*10^{18}GeV$ for the $m_{pl}$ we
get numbers like
\begin{subequations}
\begin{eqnarray}
  1.22*10^{19}GeV * 6.2*10^{-7} &=& 7.6*10^{12} GeV\\
  1.22*10^{19}GeV*1.6*10^{-7}&=& 1.9*10^{12}\\
  2.43*10^{18}GeV*6.2*10^{-7}&=& 1.5*10^{12} GeV\\
  2.43*10^{18}GeV*1.6*10^{-7}&=& 3.9*10^{11}GeV.
\end{eqnarray}
\end{subequations}
In fact the highest one of these Liddle numbers for $H_{end}$
is only a factor 7 lower than the ``unification'' scale $5*10^{13}GeV$ and
that is
not even one order of magnitude.

\item{\bf Enquist}
  
In an article by Enquist \cite{Enquist} we find
\begin{eqnarray}
  (V/\epsilon)^{1/4} &=& 0.027M\\
  \hbox{where } M &=& M_P/\sqrt{8\pi} \hbox{ is the reduced Planck scale}
\end{eqnarray}

Taking a typical (but not neccessarily true) value
\begin{subequations}
\begin{eqnarray}
  \epsilon (typical) &\sim& {\cal O}(0.01) \\
  V^{1/4} &\sim& 2.43**10^{18}GeV *0.027*\sqrt[4]{0.01}\nonumber\\
  &\sim& 2*10^{16}GeV\label{inflationV}\\
  \hbox{and Enquist says } H(\hbox{during inflation})&\sim& 10^{14} GeV
  \label{inflationH} 
  \end{eqnarray}
\end{subequations}

Enquists value is a factor 2 above the ``unification'' scale, so
we can indeed claim as our model predict that the scales of unification
and inflation should coincide order of magnitudewise. Liddle is below and
Enquist above the unification scalewith their inflation scales.

\end{itemize}
\begin{center}
  {\bf The (fourth root of) Inflation Potential $v^{1/4}$ }
  \end{center}

\begin{itemize}

\item{\bf Belleomo et al.}

In the article by Belleomo et al\cite{Bellomoetal} we find
\begin{eqnarray}
  V^{1/4} &=& \left ( \frac{3}{2}\pi^2r{\cal P}_{\zeta} \right )^{1/4}M_P
  \sim 3.3\times 10^{16} r^{1/4} GeV
\end{eqnarray}

Here $r$ is the ration of the tensor spectrum $\Delta_t(k_*)$ to the
scalar one $\Delta_{{\cal R}}(k_*)$
\begin{eqnarray}
  r&=& \frac{\Delta_t(k_*)}{\Delta_{{\cal R}}(k_*)}.\\
  r&=& 0.056\pm 0.007\\
  \hbox{giving }r^{1/4} &=&0.486\\
  \hbox{and }V^{1/4} & \sim & 1.6 *10^{16}GeV.
  \end{eqnarray}
\item{\bf Liddle}
Liddle\cite{anintr} gives a value
\begin{eqnarray}
  V^{1/4}&\approx& 10^{-3}m_{pl}\approx 10^{16}GeV
\end{eqnarray}
\end{itemize}
These scales for $V^{1/4}$ seems a bit too high for  our argument that
the highest energy density possible should be given by the lattice
scale given by the $a_0$.I.e. we should have gotten the same value for
$V^{1/4} $ in inflation as the $5*10^{13}$ which was the unification scale.

But it seems the fourth root energy density at unification were about
200 times larger than unification scale. But the Hubble Lemaitre constant at
inflation seems to match our model better.

\subsection{The Three Lowest Energy Scales}

In order to argue for the effective $q$-value for the Fermion tip scale
and two other scales (nemely string and monole), we have to first discuss
how to average in a couple of
different ways over the hypercubes in a fluctuating lattice.

One can indeed imagine a couple of ways of averaging:

\begin{itemize}
\item One way to average is simply to count all lattice hypercubes say
  equally whether they happen to be big or smallin the fluctuating lattice.
\item Another way consists in asking for what are their effect for a
  little finite or may be infinitesimal piece in space-time, because that is
  what will appear as the effect from a certain little region in space time.
  Now it will be so that if the lattice in locally dense, which means
  if $a$ is small then a higher number of hypercubes will be counted into
  a given small regon. And oppositely, when the lattice has big masks only a
  few hypercubes will be counted into a given little region. The factor by
  which we get an overcounting by asking for the contribution to a given
  little region is of courseproportional to $1/a^4$. Of course this factor
  $1/a^4$ is the ssame for the kinetic and the non-kinetic terms discussed
  for the first four ``simple'' scales, and thus this makes no
  difference for them. But it means that had we asked for an absolute
  value of a term in the Lagrangian then we should use an averaging with
  an $1/a^4$ included.

  So when we above denotedthe ``genuine'' average link size by $a_0$, it
  was an average using in fact such an $1/a^4$ weighting.
  \end{itemize}

So honestly speaking we should define two different average link sizes
\begin{subequations}
\begin{eqnarray}
  a_0 &=&\frac{ <a*1/a^4>_{counting}}{<1>_{counting}}\\
  \hbox{and } a_{0\; counting} &=& <a>_{counting}\\
  \hbox{Of course } a_{0\; counting} &=& a_0 *\exp(\sigma/2 *4) 
\end{eqnarray}
\end{subequations}
by similar ausing the Gaussian distribution as we did above.

So the true maximum in the distribution of $a$-sizes in the sense of just
counting hypercubes is thus really bigger than the size corresponding
to the ``unification scale'' $a_0$ by a factor $266^4$ =$5*10^9$.
Thinking in energy scales, the scale corresponding to the by counting
maximum in the number of links is $5*10^9$ times lower in energy than the
unification scale.

\subsubsection{The Fermion Tip Scale}

This ``Fermion Tip'' scale is defined as an extrapolation of the density on
the energy/mass axis of masses of Fermions in the Standard Model  to an
energy point,
at which the density of Fermion masses in the Standard Model  distributed on
the (logarithmic) energy axis  goes to zero. If we talk the language of
calling a Fermion
``active'' at higher energy scales $E$ than the mass, the tip-point we
talk about is the point where all the Standard Model Fermions just have become
active according to a smoothed out distribution. 

\begin{figure}
\includegraphics[scale=0.6]{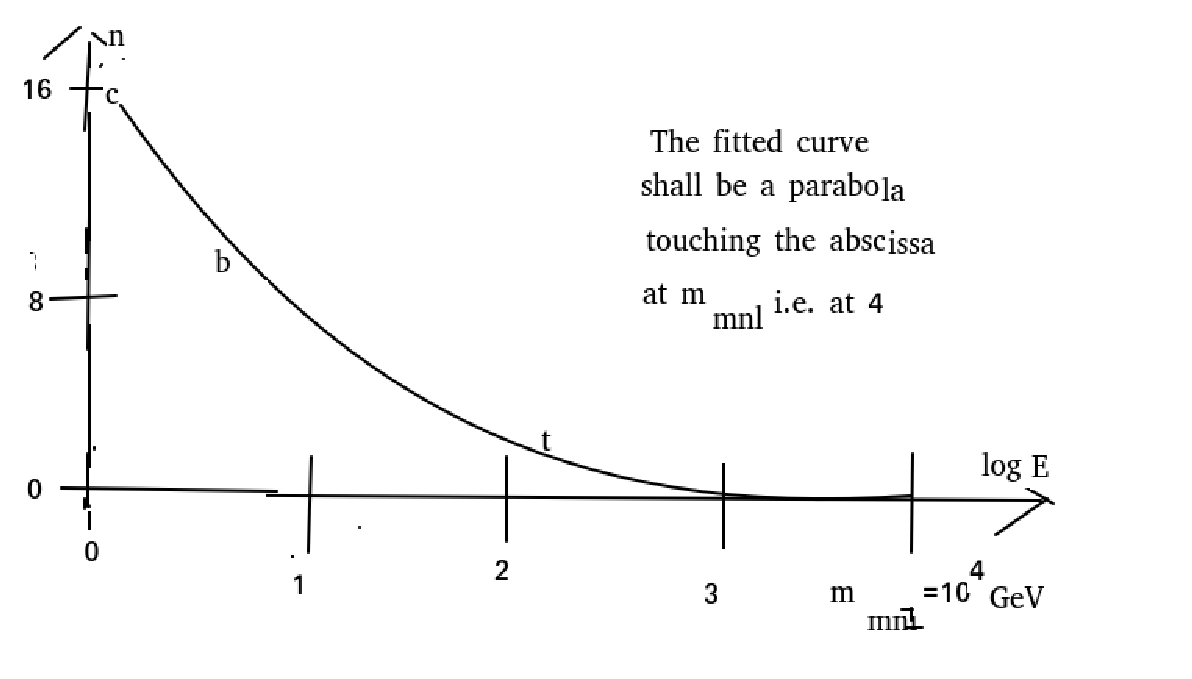}
\caption{ \label{ft} Because the distribtuion of log links after size has a
  maximum
  at the fermion tip point, we suggest that also the density of activ
  fermions(= fermions with lower mass than the scale $E$, at which you ask
  for the density of links) should behave this way, meaning with a
  {\bf parabola} behavior with maximum at the point $m_{nml}$(= the fermion
  tip point).
  If we take it that there are 45 or 44 chiral fermions per family the number
  of  families will at chiral fermion number $n$ counted from the top
  downwards in mass, be
  {``Number of active families''}=(3*45-n)/45.
  Note that the three shown point, c,b, and t, fit better on the
  parobola than on a straight line, and so the fitting form chosen is
   somewhat an empirical support.
}
\end{figure}

The philosophy of this fermion tip scale is a somewhat complicated speculative
story containing the assumption that the number of active families should
for the low energy scales below the ``fermion tip'' point equal what we can
call the thickness of the lattice. Really we have in mind a mutiplication of
the lattice so that in reallity should be one lattice for each family.
This is what would be the case if we had at higher energies not
simply the Standard Model Group but rather a cross product of one
Standard Model group for each family. On this type of model we worked
a lot and it is called AntiGUT model\cite{RDrel1,RDrel2,RDrel3,RDrel4,Don93,
  Don137}.

For the use we make of this fitting of the fermion masses for the scale
``fermion tip'' it is only important that
\begin{itemize}
\item we extrapolate to the tip whatever distribtuion may fit,
\item that this tip point reprents the maximum density of lattice hypercubes
  by simple counting (not using field theory). It is this fact that
  leads to it having
  \begin{eqnarray}
    q|_{fermion \; tip} = 4.
    \end{eqnarray}
  \end{itemize}
  
We make this extrapolation to a point, where the extrapolated Fermion mass-value
density by fitting with a formula inspired by our model of this article, the
fluctuating lattice with Gaussian distribution in the logarithm, just goes to
zero. Really our fitting curve for the mass-value-density on the logarithmic
mass axis is a parabola touching the zero-density axis at the top point just
at the point we call the ``fermion tip'' scale.  

The model behind our fitting formula is based on the physical, but very
speculative assumption:

{\bf Speculation for fitting:} At any energy scale $E$ below the
``Fermion Tip'' point the number of ``active fermions'' at that scale  is
proportional
to the number of say hyper-cubes in the fluctuating lattice with a link
size corresponding to that energy ($a^{-1}=E$ is the correspondance). Here
``active fermions'' mean fermions with mass $m < E$.

So for example for $E$ larger than the top-mass all the fermions in the
Standard Model are formally ``active''and the maximal amount of ``active''
Fermions is reached there. So the density must go to zero essentially
at the top-mass; but now we want to replace the true as function irregular
density
of fermions on the energy/mass scale axis by a more smoothed out continuous
function, and that brings the maximal point which we call $\mu_{mnl}$
somewhat higher up.

Our next assumption for the fitting function is, that 

The number of ``inactive fermions'' as function of the scale $E$ shall
grow downward in energy $E$ as {\bf the square of the distance in logarithm}
  $\ln(\frac{m_{mnl}}{E})$ i.e. as $(\ln(\frac{m_{mnl}}{E}))^2$.
      

      In order to see that such fitting hypotesis indeed are not in
      violent disgreement with the spectrum of Fermions counting their
      numbers weighted by the number of spin and color states,
      we shall present a table made for the adjust value of
      the tip-pont being
      \begin{eqnarray}
        \hbox{Tip point }&=& m_{mnl} = 10^4 GeV,
        \end{eqnarray}
      but the point is that one can see, that the quantity, that should then
      be constant from fermion to fermion is indeed very little varying.
      
      

\begin{center}
  {\bf Description of two tables for the ``fermion tip'' scale}
\end{center}

In the below tables, one for quarks and one for the charged leptons
- the seperation is quite irrelevant and we could just have put them  in just
one table -
in the Standard Model, we have first column the name of the flavor.
In column 2 we list the average for the group of particles of this
flavour (there are three colors and two spins for each flavour, making us
think of each flavour of quarks  as representing 6 =2*3 states) of their
numbers in mass
counted from above. E.g. the bunch of 6 bottom quarks have numbers
from 7 to 12 and the average is given as 9. In the third column we give
the basis 10 logarithm of the mass of the bunch of fermions with the flavour
named in GeV. In the next column, the fourth one, we give the difference
of this logarithm from that of our ansatz for the tip $m_{mnl} = 10 GeV$,
a value, which we have decided on through several attempts and drawings
of plot; I.e. we give $\log_{10}m_{mnl} = 4$ minus the logartih of the mass.
The fifth column gives this difference squared and finally the sixth column
gives this difference squared divided by the average number in the counting
from above,i.e. $diff^2/n$.  If density of mass-values along the
logarithm of the mass axis was indeed going as a constant times $diff^2$, then
the number $diff^2/n$ should be constant (actually the constant). 
Indeed the table shows, that this is approximately true.

We illustrate the type of curve to which we fit in figure \ref{ft}.

\begin{center}
  \begin{eqnarray}
    {\bf Quark\; for\; m_{mnl}}&=&{\bf 10^4GeV }\label{mnml}
    \end{eqnarray}
\end{center}
\begin{adjustbox}{width=\textwidth, center}
\begin{tabular}{|c|c|c|c|c|c|c|}
  \hline
  Name& number $n$&Mass $m$&$log_{10\; GeV}m$&$diff$=$4-log m$&
  $diff^2$&$diff^2/n$\\
  \hline
  top&3$\pm$ 1&172.76  $\pm$ 0.3 GeV&2.2374$\pm$ 0.0008&1.7626&3.1066$\pm$ 0.003&1.0355
  $\pm$ 0.001 $\pm $0.4\\
  \hline
  bottom&9$\pm$ 0.3& 4.18$\pm$ 0.0079GeV& 0.6212$\pm$0.001&
  3.3788&11.416$\pm$ 0.01&1.268$\pm$ 0.001$\pm$0.03 \\
  \hline
  charm&17 or 15&1.27$\pm$ 0.02&0.10382$\pm$ 0.009&3.8962&15.180$\pm$ 0.07&
  0.893 $\pm$ 0.004 $\pm$ 0.06\\
  \hline
  strange&25 or 23&0.095$\pm$0.006 GeV& -1.0223$\pm$0.003 &5.0223&25.223
  $\pm$0.03&1.009$\pm$0.001$\pm$0.1\\
  \hline
   down& 31&4.79$\pm$ 0.16 MeV&-2.3197$\pm$0.01 &6.3197&39.939$\pm$ 0.06&1.288
   $\pm$ 0.002 \\
   \hline
  up&37&2.01$\pm$ 0.14 MeV&-2.6968$\pm$0.03& 6.6968&44.847$\pm$ 0.4&1.212$\pm$
  0.01\\
  \hline
\end{tabular}
\end{adjustbox}
\begin{center}
  \begin{eqnarray}
    {\bf Leptons\; for\; m_{mnl}}&=&{\bf 10^4GeV }\label{mnml2}
    \end{eqnarray}
\end{center}
\begin{adjustbox}{width=\textwidth, center}
\begin{tabular}{|c|c|c|c|c|c|c|}
  \hline
  Name& number $n$&Mass $m$&$log_{10\; GeV}m$&$diff$=$4-log m$&
  $diff^2$&$diff^2/n$\\
  \hline
  $\tau$&13 or 19&1.77686$\pm$0.00012&0.2496$\pm$ 0.00003&3.7503&14.065$\pm$
  0.0003&1.082
  $\pm$ 0.00002 $\pm $0.4\\
  \hline
  mu&21 or 27&105.6583745$\pm$ $2.4*10^{-6}$MeV&$-0.9761...\pm$ $10^{-8}$&4.9761&
  24.761$\pm$ $10^{-7}$&1.179$\pm$ $4*10^{-9}$ $\pm$ 0.3\\
  \hline
  electron&41&0.51099895069$\pm$$1.6*10^{-10}$& -3.2915$\pm$ $4*10^{-10}$&
  7.2916&53.167$\pm$ $10^{-8}$&1.297$\pm$ $10^{-11}$\\
  \hline
\end{tabular}
\end{adjustbox}

Our fitting of the fermion mass spectrum is based on a philosophy that
below the energy scale $m_{nml}$ we assume a density of ``active''
(meaning  at the scale effecitvely massless) Fermion components
$45-n(\mu)$ (where $n$ is the number in the mass series counted from
high mass at the running scale $\mu$) to be proportional to
\begin{eqnarray}
  45-n(\mu) &\propto& \exp(-\frac{\ln(\mu) -\ln(m_{nml}))^2}{2\sigma_f})\\
  \hbox{or for small $n(\mu)$ } \frac{n(\mu)}{45} &\approx&
  \frac{\ln(\mu) -\ln(m_{nml}))^2}{2\sigma_f}\\
  \hbox{so that }\frac{(\ln(\mu) -\ln(m_{nml})^2}{n(\mu)} &\approx &
  \frac{2\sigma_f}{45}\\
   \hbox{or } \frac{diff^2}{n}&\approx& \frac{2\sigma_f}{45},
  \end{eqnarray}
with the notation of the table.

Taking from these  tables
\begin{eqnarray}
  \left. \frac{diff^2}{n}\right |_{quarks} &=& 1.12\\
  \left. \frac{diff^2}{n}\right |_{leptons} &=& 1.19\\
  \hbox{Take average } \frac{diff^2}{n}&=& 1.14.
\end{eqnarray}

So
\begin{eqnarray}
  1.14 &=&  \frac{2\sigma_f}{45}\\
  \hbox{and thus }\sigma_f&=&\frac{1.14*45}{2} =25.7
  \end{eqnarray}

We would have expected in our philosophy, that this spread $\sigma_f=25.7$
from the fermion distribtuion should have been the same as the spread
$\sigma = 11.0$ from the distribtuion of the scales, which is our main
interest in this article. But they seem to deviate about by a factor 2.
This is of course embassingly too much, since the $\sigma$ and $\sigma_f$ are
already logarithms and should be more accurate than that.
But the point $m_{mnl}= 10^4 GeV$ at which we get the successful fit
of the masses of the quarks and leptons is essentially an extrapolation towards
the high mass part of the sepctrum and should be well determined, even if we
do not get the theoretical success with $\sigma_f$.

\subsubsection{The Monopole Scale}
It is wellknown\cite{nonabelian} that in lattice gauge theories one
gets monopoles, if it is
possible, i.e. if the fundamental group $\pi_1(G)$ is non-trivial, as for
Standard Model
group\cite{OR, srKorfu14, srcim4} $G=S(U(2)\times U(3))$, where it is ${\bf Z}$. The Standard Model Group
in the O'Raifeartaigh\cite{OR} sense is the group with the Lie
algebra as the Standard Model $U(1)\times SU(2)\times SU(3)$ but with its
global structure arranged so that the  representations of the {\bf group}
automatically obey the quantization rules electric charge being integer for
colorless particles and have the right charge as known for the colored
particles, quarks. The group mannaging that is the group
$S(U(2)\times U(3))$, which consists of 5 by 5 matrices with the
matrices for $SU(2)$ and $SU(3)$ along the diagonal and the determinant
of the full 5 by 5 matrix restricted to be unity.

On the lattice with link variables taking values in the Standard Model Group
one will now expect Dirac strings and associated monpoles to
the elements in $\pi_1\left (S(U(2)\times U(3))\right )$, which means
closed non-contractable loops inside this group. So there should be different
types of monopoles corresponding to such loops inside the group. The  ones
with the smallest abelian monopole charge will have both weak and color
magnetica
charge, too. So we might expect the easiest combinations to get hold
of would be pairs of monopoles confined much like the quark anti quark pairs,
but now with color magnetic binding instead of color-electric fields
connecting them.

In whatever form they should be found the order of magnitude of the mass
of them should be estimated by thinking of the track of a monopole as a series
of cubes in the lattice with each of the say 6 plaquette sides carrying
approximately 1/6 the monpole charge. Each such cube in the series would cost
of order unity in the action, and so the action for a long such chain
of monopolic cubes, or simply a long time track for a monopole, will
contribute to the action a term of the order of the chain measured in
link-lengths. A similar time track in the continuum limit
is of the order $\int m ds$ where $ds$ is the infinitesimal distance
element along the time track, and $m$ is the mass of the particle, here
the monopole. So we identify
\begin{eqnarray}
  \int m ds & = & \# \hbox{monopole cubes} * O(1) /a,
  \end{eqnarray}
where $a$ is the link size of the lattice used. This of course gives us
- what must be true from dimensional arguments anyway - that
\begin{eqnarray}
  m & \approx & <1/a>,
  \end{eqnarray}
where it is easily seen that the averaging taken in the pure counting sense,
since we did not use any field theory here. This means an effective $q$
to get the average must be
\begin{eqnarray}
  q_{eff}\hbox{(for monopoles)} &=& 4 +1 =5.\label{q5}
\end{eqnarray}

This sugests from our fitting with the straight line an energy scale
of the order $10^4 GeV/(250 \; or \; 266) =$ 40GeV or 38GeV.
There has been seen extremely little deviations from the Standard Model in
LHC, but in fact there was one possible statistical fluctuattion
giving a dimuon resonance with mass $28.3\pm 0.4 GeV$ and width
$\Gamma_{\mu\mu} = 1.9\pm 1.3 GeV$
\cite{dimuon}. The statistical significance globally is 3.0 s.d..

In addition Arno Heister \cite{Heister} has by analysing old ALEPH
data found an enhancement at 30 GeV in mass for dimuon.

We may take these uncertain observation as seeing a monopole scale at
\begin{eqnarray}
\hbox{``The monopole scale'' }&=&28 Gev.\label{monopole}
\end{eqnarray}

\subsubsection{The String Scale}
Let us think, that there were some string theory scale and we know the usual
Nambu-Goto-action being of the form
\begin{eqnarray}
  \hbox{Nambu-Goto action } {\cal S} &=& -\frac{T_0}{c}\int d{\cal A},
  \end{eqnarray}
where the integral over the area measure $d{\cal A}$ means the area of the
time-track of the string. The coefficient is in partly historical  notation
\begin{eqnarray}
  \frac{T_0}{c} &=& \frac{1}{2\pi \alpha'},
  \end{eqnarray}
where $\alpha'$ is the Regge pole slope used in Veneziano model.

It is not surprising, that an area proportional action should be proportional
to the lattice link  length squared, as is also enforced by the dimensionality
of the coefficient
\begin{eqnarray}
  [T_0/c] = [1(2\pi \alpha')] &=& [area^{-2}] = [energy^2].
\end{eqnarray}
Thus the effective $q$ for the string scale,if there were one, would be
\begin{eqnarray}
  q_{eff}(strings) &=& 4+2 =6 \label{q6}.
\end{eqnarray}

An immediate estimate using say 250 or 266 for the per step in $q$
factor in the energy scale, we find, that the string energy scale
should be
\begin{eqnarray}
  \hbox{string energy scale }&\approx& 10^4 GeV/(250^2) \hbox{ or }266^2)\\
  &=& 0.16GeV \hbox{ or } 0.14 GeV\\
  \hbox{agreeing with }\sqrt{1/(2\pi \alpha')}&=&
  \sqrt{1/(2 \pi 1.1GeV^{-2})  }\\
  &=& 0.38 GeV \hbox{(order-of-magnitudewise)}
\end{eqnarray}

It suggests very strongly that we should like to take the appearance
of string-like physics in hadron physics as giving us a {\bf string scale}.
We can say with
\begin{eqnarray}
  \hbox{From $\alpha'$ } \hbox{``string scale''}&=& 0.38 GeV\label{alphaprim}\\
  \hbox{From Hagedorn temperature } \hbox{``string scale''} &=& 0.17GeV
  \label{Hagedorn}.
\end{eqnarray}
\subsubsection{Fit of the Low Energy scales}

Three of our energy scales get in the fluctuating lattice model their
average made in the simple counting hypercubes way (rather than as we do for
the field theory related scales), so here we present a fitting line for
these three
alone.

\includegraphics[scale=0.7]{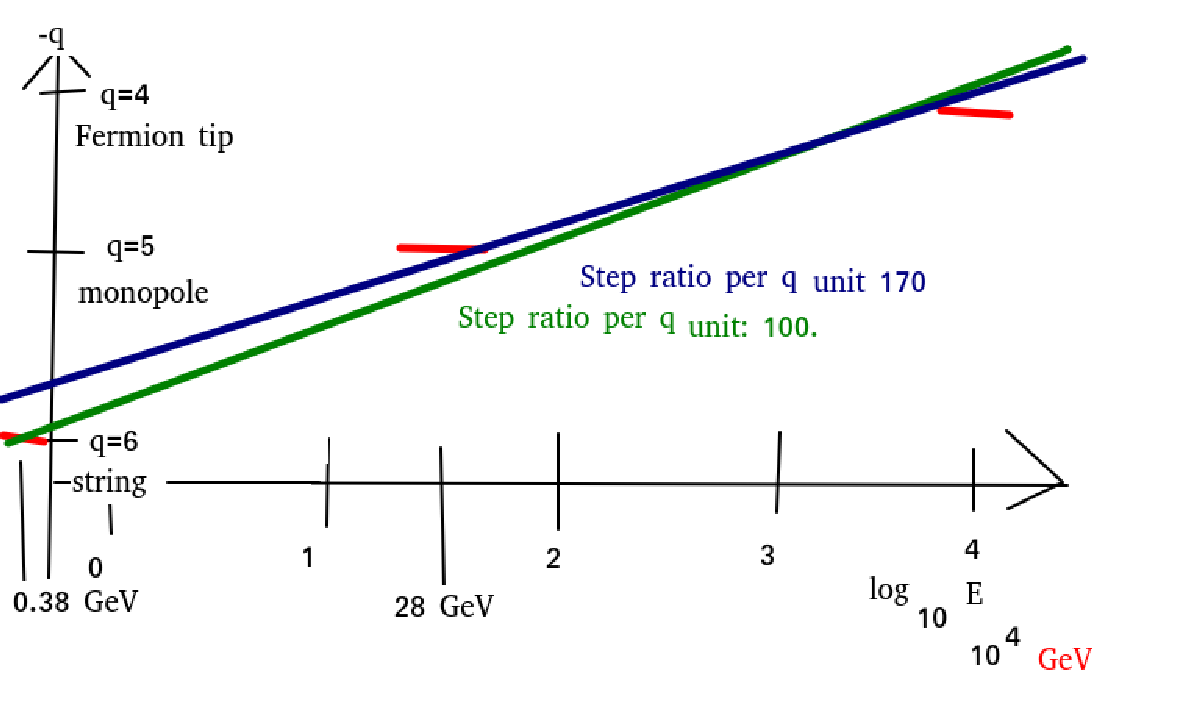}

\subsection{Total fit line}

\includegraphics[scale=0.8]{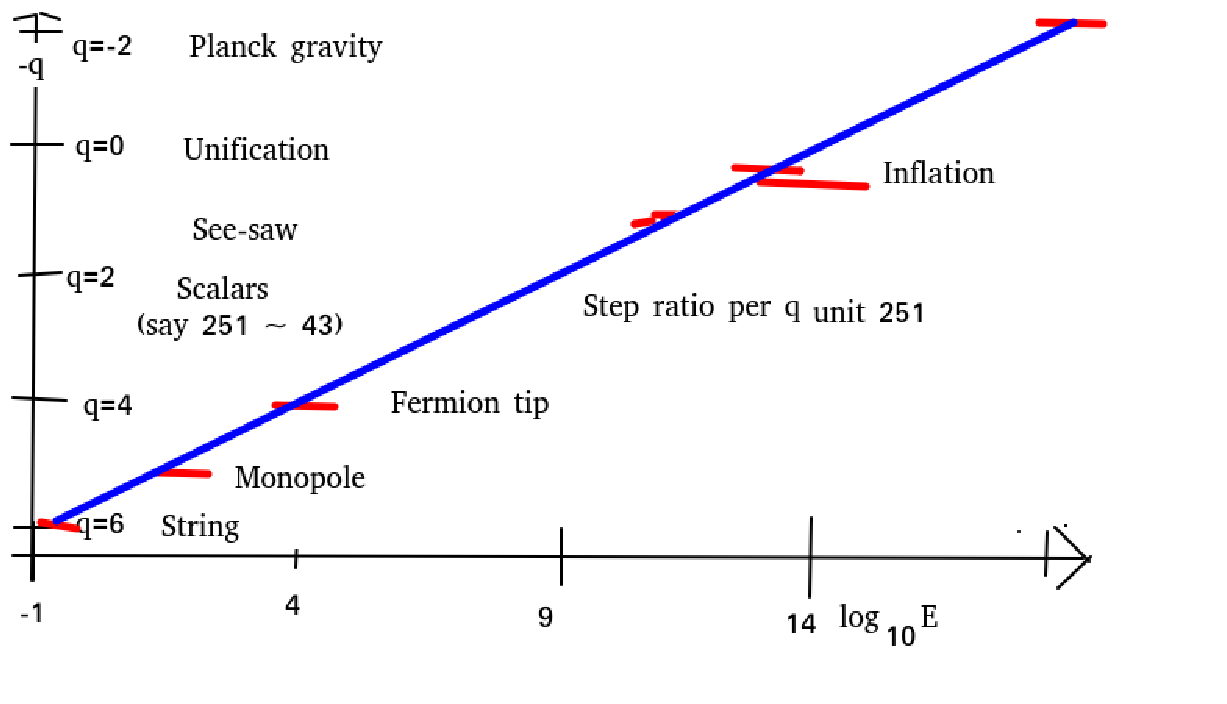}
\section{Fluctuating Gauge Argument for Fluctuating Lattice}
\label{FGA}
In the work ``Dynamical Stability...Light from Chaos'' we, F{\o}rster, Ninomiya
and myself \cite{FNN, RS}, suggested, that gauge theories could appear with
actually exact gauge symmetry, even if fundamentally there was only such
an approximate symmetry. The reason should be, that very strong quantum
fluctuations  occur in the gauge, even if the gauge symmetry is weakly
broken in the fundamental action. Thus the last small breaking is washed out.
Note that in our philosophy there exists a priori fundamental variable
corresponding to the gauge degree of freedom, although it might be interpreted
differently in pracsis. But there is ontologically existing d.o.f. comming
from that. For the gravity case M. Lehto et al\cite{trans} argued for deriving
translational invariance in this spirit.
If this
is the mechanism for getting also the gauge symmetry of reparametrization
in gravity, then the coordinate system will fluctuate strongly. Here we  have
the point of view that some gauge degrees of freedom exist in Nature, it is
ontological. Now such fluctuations in coordinates is supposed to carry a
lattice with it, and we who work with a coordinate system, which is sepcified
by some conditions, will think of a cooordinate system very different from
the ontological one. We would rather put the story, that the ontological
coordinate system and the associated lattice fluctuates wildly. That is to say
a model for obtaining gauge symmetries by quantum fluctuations actuality leads
to effectively a fluctuating lattice.

In first approximation our fluctuating lattice actually behaves just as it
should in such a philosophy for how gauge symmetry comes about, but were it
taken completely serious the fluctuations of the lattice should fluctuate
soas to in some superposition components divergently deviating from others.
In our fit we have a convergent fluctuation in the link scale say; but it
is large in the sense that it gives rise to large numbers,our 251 steps.

Some somehow there is a need for making the fluctuations of the lattice
limited in stead of the a priori infinite scale fluctuation, that would
correspond to perfect reparametrization symmetry.

I would say, that having a large number, our 251, in the place whee it should
a priori have been infinite is somewhat nice.
\section{Conclusion}
\label{con}
We have in this article connected a series of energy scales found in
physics - and strictly speaking expected to be equal to
``{\bf the}  fundamental energy scale''. But the 8 scales, we have considered
are deviating significantly even in order of magnitude. This is philosophically
somewhat unexpected. The present article has repaired on that problem by
proposing, that a fundamentally existing {\bf fluctuating} lattice could unite
these 8 energy scales as all comming from the same fluctuating lattice. They
are thus all deduced from the distribution of the link size for the fluctuating
lattice. In our simple Gaussian in the logarithm ansatz this means that the
eight energy scales are given by - fitted by - just two parametrs, the overall
scale $1/a_0$ and a width of the Gaussian distribution assumed  - which is
dimensionless -, this width being
$\sigma \sim 11$. Instead of the width we might represent the parameter
which so many energy scales can be fitted as the step factor by which
the energy scale goes each time a quantity $q$ related to the dimensionality
of the coefficients of the actions involved in the energy scales is lowered
by one unit. This then one could say ``holy parameter'', the step factor
is about 251.

If one would take the scales with large negative $q$-values seriously, then
we have at least some of the fundamental scales in the region, where present day
and near future experiments have a chance to find e.g. non-locallity
or other effects of the lattice. This gives much more
promissing aspects for investigating fundamental physics by
experiments in not so far future, than if - as I believed myself
untill recently - the Planck scale was the fundamental scale for
physics.

With this work we would rather consider e.g. the ``fermion tip scale''
the most fundamental one. And the fermion tip scale is already
formally reached by LHC, although to really see, say, if there were
non-local effects at $10^4$ GeV might require in practice somewhat higher
energy accelerators.

\subsection{Failiors}
At the end we should still admit, that although we even have invented
a ``scalars'' scale, with the mass of which there should exist a lot
of scalar bosons, then the by now known scalar, the Standard Model Higgs, has
a so fine tuned mass
that it does not at all fit in to the ``scalars scale''. Personally I would
expect, that to get the Higgs mass sufficiently fine tuned something like
our complex action theory in which fine tunings can be modelled and
even can depend on the future\cite{CAT, CAT2, Coincidensies, Relation}.

Similarly one might have hoped for obtaining the cosmological constant
by a formalism analogous to the string theory giving the
hadron string scale and the monopoles scale. One should understanding
the cosmological constant as the action density for
a space filling 3-brane, expect the value $q=4+4=8$ to deliver the
scale of the cosmological constant; but alas
\begin{eqnarray}
  \hbox{Our prediction} 10^4 GeV/ 251^4 &=& 4*10^{-7}GeV\\
  \hbox{while Cosmological const. scale }\sqrt[4]{\rho_{\Lambda}}
  &=& \sqrt[4]{10^{-47}GeV^4}\\
  &=& 1.8*10^{-12}GeV \ne 4*10^{-7}GeV\nonumber
\end{eqnarray}
(we used vacuum energy density or approximately the critical
energy density $10^{-47}GeV^4$, see e.g.\cite{Cc}). The two
numbers $1.8*10^{-12}GeV$ and $4*10^{-7} GeV$ are not even order of magnitudewise
close. So the cosmological constant also fail in our model and is somehow
fine tuned!(a finetuning completely spoiling our prediction in the present
article) Personally I woud say that a finetuning has overwritten the
our order of magnitude model\cite{Frampton},  In this work we give an
argument for zero or cosmologicalconstant in the complex action and influence from future spirit.

\subsection{Yet a possible scale?}
However, looking similarly for domaine walls or 2-brains we would get the
predicted scale $10^4 GeV/251^3 =6*10^{-4}GeV= 0.6 MeV$, which is not far from
the energy scale for the tension in the domaine walls fitted
to Colin Froggatts and mine dark matter model, in which we have a scale
of the order of a few MeVs (say the wall tension we fitted
$S=(8MeV)^3$.\cite{odm}.
\section*{Acknowledgement}
The author thank the Niels Bohr Institute for status as emeritus.
This work was discussed in both Bled Workshop and the Corfu Insitute
this year 2024.


\begin{thebibliography}{90}
\bibitem{Stojkovic}Dejan Stojkovic ``Can we push the fundamental Planck
  scale above 1019 GeV?'' htttps://cds.cern.ch
\end{thebibliography}
\end{document}